# SOUNDBAY: DEEP LEARNING FRAMEWORK FOR MARINE MAMMALS AND BIOACOUSTIC RESEARCH

**Noam Bressler**
Deep Voice
noam@deepvoicefoundation.com,

**Michael Faran**
Deep Voice
michael@deepvoicefoundation.com

**Amit Galor**
Deep Voice
amit@deepvoicefoundation.com

**Michael Moshe Michelashvili**
Deep Voice
moshe@deepvoicefoundation.com

**Tomer Nachshon**
Deep Voice
tomer@deepvoicefoundation.com

**Noa Weiss**
Deep Voice
noa@deepvoicefoundation.com

## ABSTRACT

This paper presents Soundbay, an open-source Python framework that allows bio-acoustics and machine learning researchers to implement and utilize deep learning-based algorithms for acoustic audio analysis. Soundbay provides an easy and intuitive platform for applying existing models on one's data or creating new models effortlessly. One of the main advantages of the framework is the capability to compare baselines on different benchmarks, a crucial part of emerging research and development related to the usage of deep-learning algorithms for animal call analysis. We demonstrate this by providing a benchmark for cetacean call detection on multiple datasets. The framework is publicly accessible via https://github.com/deep-voice/soundbay

## 1 Introduction

Bioacoustics research relies heavily on the analysis of acoustic data. Passive acoustic monitoring systems have been in ongoing use since the 1970s, and have become a research and industry standard for gathering audio signals in a soundscape. The improvement in telecommunication and electronic storage leads to the accumulation of enormous amounts of acoustic data [1]. Given the large amounts of acoustic data available and needed in the bio-acoustics research field, automation, and data handling methods are required. Significant tasks that the scientific community is focused on include the detection and classification of acoustic events [2, 3] [1]. Those instruments can be achieved either by using classic signal-processing algorithms, data-driven solutions, or a combination of both. Nowadays, deep learning methods are the best-performing algorithms for classification and detection in many domains, such as image, text, and audio [4, 5, 6]. The emergence of deep learning is also noticeable in the field of bio-acoustics [7, 8, 9], and used for the calls analysis of birds [10, 11, 12, 13, 14, 15], anurans [16, 17, 18], insects [19, 20, 21], fish [22, 23, 24] , terrestrial mammals [25, 26, 27, 28, 29] and marine mammals [30, 31, 32, 15, 33]. While some models have been reviewed and developed for analyzing marine mammals' vocalizations [30, 34, 35, 36], there are still few benchmark works that test the effectiveness of those models in general [37]. Nevertheless, comparing the different methods can be beneficial when evaluating the models' performance. Such a comparison can assist in choosing a specific deep-learning model for a particular dataset or task at hand. Another reason for the need for standardization is the overhead of creating a new deep-learning model using existing tools. Most open-source implementations for deep learning algorithms for audio detection and classification are very specific: they use a single model and usually provide an ad-hoc method for data pre-processing. Using those packages for new domains, datasets, or models is a complex task that requires high expertise and code re-organization. Some other repositories try to tackle the standardization problem [38, 39]. Here, we provide an end-to-end solution for the two significant issues mentioned. We create a benchmark and baseline comparison between a few deep learning models in the field of acoustic detection algorithms. We demonstrate it on cetacean calls datasets. The framework has no assumptions about the recording apparatus or acoustic environment. Hence, it's relevant to any animal vocalizations. Soundbay provides easy and modular creation of deep learning algorithms with little to no code needed. Soundbay also provides some unique features such as architecture search and hyper-parameters optimization integrated into our framework [40]. Also, importing new models from the community is easily done in the framework suggested. By introducing Soundbay, we plan to serve as a baseline and go-to framework in bio-acoustics detection and classification.

---

[1] https://www.soest.hawaii.edu/ore/dclde/

Soundbay serves as the deep learning engine in the automatic annotation, detection, and classification services provided by Deep Voice. The service includes a data platform where biologists and environmentalists can upload their data, train models, and receive annotated results on their respective data. More details can be found on our website [2]. Contact Deep Voice directly for any further assistance or information [3].

## 2 Soundbay Framework

Soundbay breaks down the process of building an acoustic detection and classification models into several components, where each component is independent and composed of interchangeable building blocks, including their hyper-parameters. The components are:

- Model - The type and architecture of the trained model.
- Optimizer - The type of optimizer and loss function used to train the model.
- Data - Handling the data on which the model is trained on. This includes features such as training and validation splitting, distribution equalization, sampling, and data loading.

The subsequent two sub-modules are part of the data component and attain their hierarchy following their variability in different papers [41, 42] and importance on the output result:

- Pre-processing -Transforms applied to the acoustic raw signal before being used by the model for training or inference, usually moving to a different audio representation.
- Augmentations - Operations are applied to the training data to generate variations on the original signal - thus increasing model generalization.

Soundbay enables the framework user to automatically explore the space of possible building blocks and hyper-parameters. A new experiment can be created by changing only a single component, as the components are not coupled. An essential feature of this design is hyper-parameters building-blocks optimization. This enables inspection of different types of blocks for a specific problem with little boilerplate (see section 2.1.5). Soundbay's deep learning modules and automatic differentiation operations are based on the popular PyTorch [43] framework.

In subsection 2.1 we describe the main features and capabilities of the framework. In subsection 2.2 we describe the structure of the framework and further describe each component in the pipeline.

### 2.1 Main features

#### 2.1.1 Model Training

Each experiment is carried out by a single command line, specifying the path to a configuration file that contains all the hyperparameters needed to run an experiment. Changing hyper-parameters doesn't necessarily imply the writing need of a new configuration file, since overrides are supported by the Hydra package mentioned in 2.2.1. An example of a training command is found below:

```
python train.py --config-name runs/main_unit_norm data.batch_size=8
    optim=jasco_vgg_19 experiment.manual_seed=4321
```

---

[2]https://www.deepvoicefoundation.com
[3]info@deepvoicefoundation.com

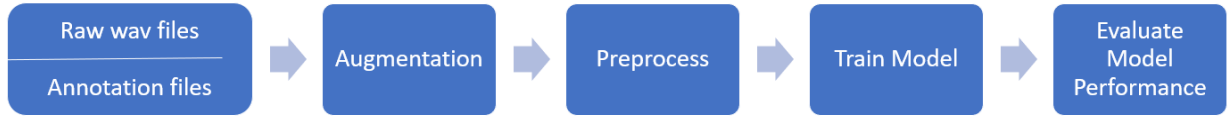

Figure 1: Training process of a model. Samples are loaded from the raw wav files, and labeled based on the annotations. Afterward, augmentation and preprocessing functions are applied to the samples. Then, the samples are fed into the model, and the prediction loss is calculated using the ground truth labels. During the model training, a validation process occurs on the validation data to test the model's performance on data it didn't train on. Predictions for new data can be made at the end of the training session.

The command runs the experiment in the file `conf/runs/main_unit_norm.yaml`, which serves as the global configuration file. Using command line overrides, the parameters for batch_size and manual_seed are amended, and we choose the group-level optimization configuration specified in `conf/optim/jasco_vgg_19.yaml` instead of the one in the global configuration file. A detailed explanation of the configuration files can be found in 2.2.1. During training, the model is evaluated using a validation subset. The loss and relevant performance metrics are logged and uploaded to the logger.

### 2.1.2 Finetune

Finetuning of a model is a common practice while training deep-learning algorithms. In this method, we take advantage of a pre-trained model and use it as a starting point for training, instead of randomly initializing the weights of a new model. This method is preferable when we have a small dataset in hand, or when the distribution of the new training data is similar to the distribution the pretrained model was trained on. Generally speaking, there are two options for fine-tuning a model: training all the model weights, or solely change the final weights of the model [44]. Both options are supported in our framework.

### 2.1.3 Inference

Inference and using models for decision-making are the most important features of a trained model. We support various types of prediction/inference setups:

- Evaluation of a trained model on its associated test set.
- Inference on arbitrary audio recordings based on previous learning processes. If annotated metadata are supplied, classification metrics are also returned, indicating the model performance on these new data.

The output for both options is a CSV file with class predictions and their associated probabilities over time, where the time resolution corresponds to the model training data. We provide a visualization utility for visual inspection of the detection algorithm results in the form of class probabilities and prediction scores, both displayed on the top of the audio spectrogram, Fig. 2. We also provide an annotation file compatible with Raven Lite or Pro. [45]

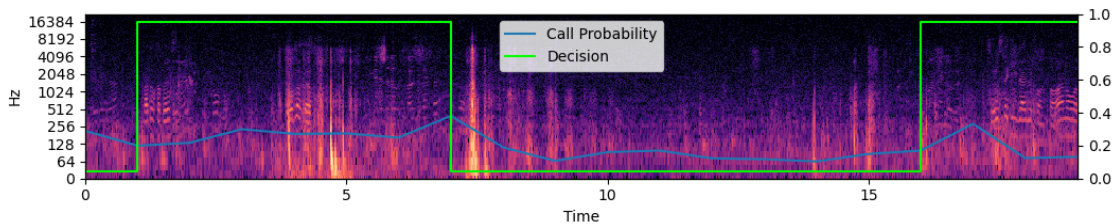

Figure 2: Inference visualization using the provided visualization tools. Recording date: 14.9.2018 at 15:01:27 PM [CAT]. The green line depicts the detection decision, while the blue line represents the raw class probabilities. The detection decision threshold is usually chosen using the operating point on the ROC (receiver operating characteristic) curve.

### 2.1.4 Experiment tracking

Experiment tracking is crucial to any research project, especially when training ML models, where experimentation with different hyper-parameters and models can be done rapidly. For that purpose, Soundbay is built in with a logging mechanism, which by default uses the popular experiment tracking online platform of Weights and Biases (W&B) [40]. When using W&B, which offers a free service for individuals, each training program run with specific settings and parameters is tracked as an "experiment" on the online platform. Each record contains all hyper-parameters used, as they are set in the configuration files and common classification metrics throughout the model training. Default tracked metrics include precision, recall, F1, AUC (area under the curve), and accuracy [46]. Tracking these metrics throughout the training process yields corresponding significant consequences (for example, whether the model is underfitting or overfitting). Inspecting those recorded metrics after the training process is used to compare different experiments using different parameters. Examples are available in the experimental paragraphs, 3.2 and 3.3.

### 2.1.5 Hyper-parameter optimization and sweeps

Soundbay enables using the "sweep" mechanism developed by Weights and Biases [40]. This feature enables the definition of a particular search space of possible hyper-parameters in which the best possible configuration is assumed to be found. The search is usually conducted to optimize one or more metrics, for example, finding the setup with the higher F1 score. We can either scan the whole space, combination by combination (also called grid search [47]) or use a dedicated hyper-parameter optimization algorithm that will attempt to find the combination that optimizes a given metric in a computationally efficient way [48]. The results from all the experiments performed within a sweep are saved to the W&B platform in the Sweeps section. An example for the usage of a sweep in the framework is given on 3.2.

## 2.2 Framework structure

Here, we present the current building blocks of the framework and the implemented classes available. It's worth mentioning that custom-made classes are also supported in the framework and thus their usage is not restricted to the following components.

### 2.2.1 Configuration files

One of the goals of the Soundbay package is to accelerate the development of new models. When training ML models, a core part of a researcher's work is experimentation with various models, pre-processors, datasets, augmentations and optimizers. In order to streamline the process of modifying these components, Soundbay uses the Hydra package [49]. This enables controlling the contents of these components in a more flexible and user-friendly fashion. Using Hydra, each component's contents are controlled by configuration files, which can determine either parameter values (for example, learning rate) or the actual object used (for example, what model class is used). The main configuration includes different parameter groups, as can be seen in the example of the default configuration:

```
# @package _global_
defaults:
  - data: defaults
  - augmentations: _augmentations
  - preprocessors: _preprocessors
  - model: defaults
  - optim: defaults
  - experiment: defaults
```

As an example for a group configuration, we can view the content of the default optim group:

```
optim:
  epochs: 100
  optimizer:
    _target_: torch.optim.Adam
    lr: 0.001
  scheduler:
    _target_: torch.optim.lr_scheduler.ExponentialLR
    gamma: 0.995
```

This optimizer configuration file contains the parameters that have to do with the optimization process. The first argument of the file sets the number of epochs, the second defines the optimizer object and its own parameters. The third argument of the file defines the scheduler object that will be used. The global-level configuration file is composed of a list of all the component configuration files. Hence, creating a new experiment can be done by creating a new global configuration file, or by using Hydra override options which are detailed on their website [49].

### 2.2.2 Data handling

The data consists of handled dataset objects: `ClassifierDataset` and `NoBackGroundDataset`, which are in charge of loading and handling the data for the model. The dataset objects expect two inputs: The path to the audio files folder and a metadata file with labels for call classes and their duration. The first dataset object assumes the gaps between calls should be used as background noise labels, while the second one prevents this option, and can be useful in cases where only differentiation between given calls is needed. The base dataset class accepts a few arguments that control the data representation, that might need to be adapted for optimal results. They include nfft, window size, normalization methods, etc. Optional features include dataset balancing for underrepresented call types and re-sampling the data to a different sample rate [50]. After loading the samples, augmentations and pre-process operators are applied to them before building the training batches. During inference, the dataset object `InferenceDataset` carries out the I/O operations needed to load the data.

**Augmentations** Using data augmentations is a common practice while training machine learning models [51]. They create robust models which are less prone to overfitting on the training data. They assist in extracting meaningful features from the audio files and might be crucial for generalizations on different devices and environments that might differ from the training subset. Augmentations are functions applied on the input signal, that alter the signal itself, but doesn't change the label of the signal. For example, if we mask a segment of a whale call using the masking augmentation, it doesn't change the call class, and the model should be unaffected by this change. In our repository, there are a few augmentations implemented that could be stacked together:

- Additive Gaussian noise - By using `torch.randn`, a random vector with normal distribution is created. The noise variance is then normalized by the input signal variance to yield a meaningful signal-to-noise ratio. Finally, the noise samples are added to the data samples.
- Temporal masking - choosing a random time frame in the given waveform and masking it as presented by Park et al [41]
- Frequency masking - choosing a random frequency range and masking it from the signal's frequency spectrum, as presented by Park et al [41]
- Chained augmentation -Each data sample is augmented randomly with either of the above-described augmentations, where each augmentation probability is given a priory as one of the model hyper-parameters.

The augmentations are implemented using sox_effect [52] from the torchaudio [53] package.

**Pre-Processing** Albeit some new architectures work on raw audio representations [54], we use a pre-processing operator that transforms an input waveform file into a spectral-temporal representation, which is the common representation for training classification models.

We include two types of pre-processors in our framework:

- Audio Representation - Changing the audio representation to include the frequency domain. This includes operations such as `Spectrogram`, `AmplitudeToDB` or `MelScale`
- Normalization - Applying the data normalization method to create a unified data representation for the model. Might include normalizations such as `PeakNormalization` (normalizing the input to a range [-1,1]) or `UnitNormalization` (normalizing the input by subtracting its mean and dividing by its standard deviation).

### 2.2.3 Models

We provide a few common bio-acoustics audio-classification models in the framework. All the models are based on convolutional neural networks. Supported models include:

- `ResNet1Channel`, `GoogleResNet50` [55], `OrcaLabResNet18` [30]

ResNet is a neural network architecture created by Shawn et al[56]. It consists of residual blocks that allow better gradient flow in the model. ResNet is a common baseline network for classification purposes.

- `VGG1Channel`, `VGG19Jasco` [36]

The VGG architecture [57] is a robust feature extractor for deep learning models relying on the foundations made by the original AlexNet [58] architecture. It has several features such as 1x1 convolutional layers and smaller receptive fields (3x3 unlike 11x11 in AlexNet).

- `ChristophCNN`, `GenericClassifier`, `ChristophCNNwithPCEN`

The mentioned models are taken from the work of Chin [59] and the corresponding repository, an open-source repository for whale sounds classification.

Aside from models, the following building blocks are available for integration and creating different "flavored" models:

- `PCEN`

The PCEN (Per Channel Energy Normalization) is a trainable network component that has been shown to improve results in audio classification and ASR (Automatic Speech Recognition) tasks [60]. Its parameters can be used with constant weights or be trained as part of the entire training process.

### 2.2.4 Optimizers

Soundbay framework uses the PyTorch [43] machine learning framework, hence all PyTorch components are easy to use within the framework, including existing PyTorch optimizers. As can be found in our configuration files, we used mainly the ADAM [3] and SGD (Stochastic Gradient Descent) [61] optimizers. Schedulers are added on top of the optimization process.

## 3 Experiments

To demonstrate the capabilities of Soundbay, two experiments were performed.

- Hyper-parameters and architecture optimization sweep, searching for the best experimental setup for a specific dataset.
- Benchmark comparison of marine mammals calls detection capabilities, inspecting various models on several whale calls datasets.

Both experiments present the ease of performing a combination of multiple runs using the Soundbay package. Appendix 6A presents the commands replicating those experiments. Section 3.1 describes the datasets used for the benchmark creation. Section 3.2 presents a case study using the framework for finding the best training setup for a given dataset. Section 3.3 presents a benchmark experiment for the datasets and models mentioned in this paper.

### 3.1 Datasets

We used three datasets for the benchmark comparison. For each dataset, we provide the information for the data processing, train/val/test splits, and some technical data about the files and call characteristics. The size for each dataset is given in Table 1.

|  | Train | | Val | | Test | |
| --- | --- | --- | --- | --- | --- | --- |
| dataset | Calls | BG | Calls | BG | Calls | BG |
| OrcaData | 1216/0.676 | 1060/1.125 | 132/0.077 | 101/0.134 | 90/0.049 | 91/0.2 |
| ICML2013 | 3253/1.803 | 22618/12.506 | 1244/0.690 | 9094/5.032 | 648/0.360 | 7355/4.078 |
| DeepVoice 2018 | 1580/0.719 | 1009/0.232 | 758/0.205 | 770/0.246 | 495/0.154 | 595/0.125 |

Table 1: Total amount of samples per dataset and the total duration, for each split and sample type. the format is #samples/duration (hrs).

### 3.1.1 OrcaData Killer Whale Calls

The Orca detection dataset is taken from the OrcaData [62] repository and includes recordings from various sessions, annotated as 'podcast' in the metadata attached to the audio files. [4]. We applied some processing on the data:

1. We removed podcast 1, which was mainly composed of old recordings. We found that the annotations for these parts weren't accurate and, hence, might degrade the performance of ML models trained on non-accurate annotations.
2. We filtered out calls of exactly 2.45 seconds. As mentioned in the OrcaData repo, those annotations are usually based on model predictions without human annotation.
3. We resampled the data to a frequency of 20000 Hz to standardize the dataset.
4. We applied train/val split to a ratio of 0.9/0.1. We separated the sets by file names so we wouldn't have recordings from the same file on different datasets. To evaluate the models on test data, we used the OrcasoundLab09272017_Test.tar.gz testset provided by the OrcaData repository.

### 3.1.2 ICML Right Whale Calls

The Right Whale Redux Challenge dataset, a Kaggle competition created as part of ICML2013 workshop [63], was selected as a benchmark dataset. The given challenge was detecting North Atlantic Right Whales within passive ship-borne array recordings. The data was split into training, validation, and test data in the following manner - recordings from 28.03.2008 were used for testing, the recordings from 30.03.2008 were used as validation data, and the rest of the recordings from 30.03.2008 and 31.03.2008 were used as training data. The sample rate of the files is 2000 Hz.

### 3.1.3 Mozambique Dataset

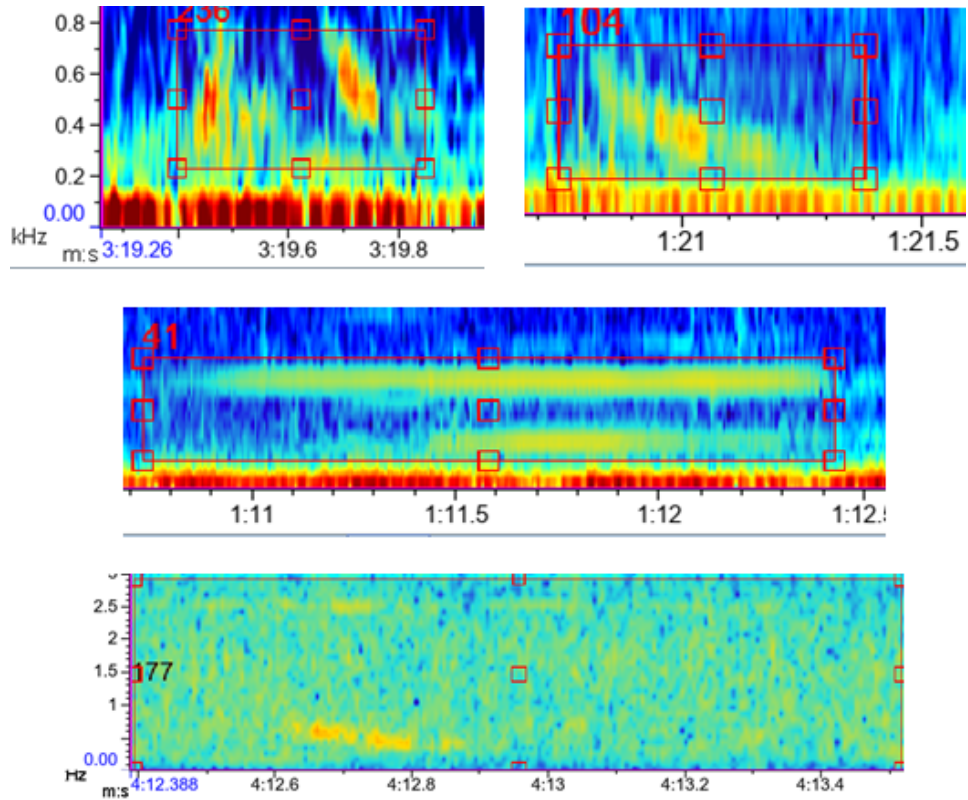

Figure 3: A few examples of the Humpback whale complex calls, taken from the Mozambique 2018 recordings. We can see an up-down sweep, a down sweep, a cry, and a grunt [64]. The red rectangles are the bounding boxes annotated by our team

---

[4]can be found in aws –no-sign-request s3 ls acoustic-sandbox/labeled-data/detection/train/TrainDataLatest_PodCastAllRounds_123567910.tar.gz

In September 2018, Deep Voice set out on an acoustics data acquisition expedition to the Bazaruto archipelago in Mozambique. This area is a breeding zone for the C1 population of Humpback Whales, which start to arrive in June and leave by the end of October [65]. Humpback whales have complex vocalizations [64, 66], varying in frequency range, tempo, and length. The three main call types are whale songs, social calls, and foraging calls. Deep Voice recorded mainly whale songs by inserting the hydrophone in randomly sampled geospatial locations. Alongside the whales' vocalizations, the Bazaruto archipelago has three species of dolphins [67] who vocalize frequently. The recordings also contain shipping noises (which are not frequent in the area), coral riff noises, background noise originating from flow noise, random turbulence, and equipment interaction with the medium. HTI-96 hydrophone connected to a zoom H5n recorder were used as the data acquisition equipment. The audio files are single-channel recordings with a sampling rate of 44.1 KHz. In total, there are 50 recordings, lasting about 12 hours. The recordings were made while the boat was stopped and the engines were off. Each recording session is accompanied by a log in the logbook, including environmental metadata, whale surfacing documentation, recording gain, and geospatial location. Each embarking to the ocean from the research station was accompanied by telemetry recording using the Delphis [68] application, which includes data regarding GPS location, gain, sea state, and surface behavior of the Humpback Whales.

**Data annotation** To train detection models on the collected data, our research team and volunteers annotated a subset of the calls from the dataset. The annotation process was conducted using Raven Lite 2 program [45]. The calls are annotated as either whale songs, whale social calls, dolphin calls, or unknown noises. Only calls associated with whales are considered as the positive class of the detection task, while the rest are considered as background noise/negative class. Bounding boxes of the annotations can be viewed in FIG.3 The annotated files were split into three different splits by dates so that data contamination wouldn't occur between samples. Recordings from 12.09.18-16.09.18 were used as train split, 19.09.18 were used as val split, and 17.09.18 were used as test split.

### 3.2 Hyper-parameters Optimization run

In this section, we inspect one of the prominent features of our framework: hyperparameters optimization and architecture search. It's worth noting that we provide an "expanded" hyperparameters search, since we're not only optimizing the parameters of a single model but also optimizing different models and their building blocks in the process. The ICML2013 dataset described above was used as the target dataset, so we would like to find the best-performing model. The experiment's goal was to find an optimal configuration for call detection, given the dataset, from a predetermined configuration space. The configuration space was built from the following options:

1. `ResNet1Channel` and `ChristophCNN` models
2. Optimizers - Adam, Christopher Chin SGD [59], and Jasco SGD [36]
3. Preprocessing - peak normalization, unit normalization, sliding window normalization
4. Augmentation probability: 0-1

We tried two different models, three different optimization schemes, three different preprocessing options, and a uniform variable for the augmentations probability. The full configuration for running the sweep can the found in the appendix 6.1.

The configuration for the optimization run contains the following details:

- Optimized goal: maximize AUC
- Model optimization method: hyperband Bayes optimization [40]
- Epochs: 100
- Early stopping: 10 epochs

The peak AUC achieved was 0.986, using the following configuration:

- Model - ResNet1Channel
- Model optimization method - Christopher Chin SGD
- Peak Normalization
- Augmentation probability: 0.821

Two significant outputs of the sweep are the parameter importance and the ROC (receiver operating characteristic) graph. While the mentioned configuration achieved the best performance, other different configurations achieved similar results, as can be viewed in the ROC graph. Further inspection of the parameter importance figure can teach us about the features that were crucial for the good performance of the model. For example, using a ResNet50 model has a high correlation with higher AUC, while using Cristoph CNN has a strong negative correlation with AUC. Using this sweep feature can help a researcher to design the best setup for a given dataset, and highlight the features or parameters that are important to use.

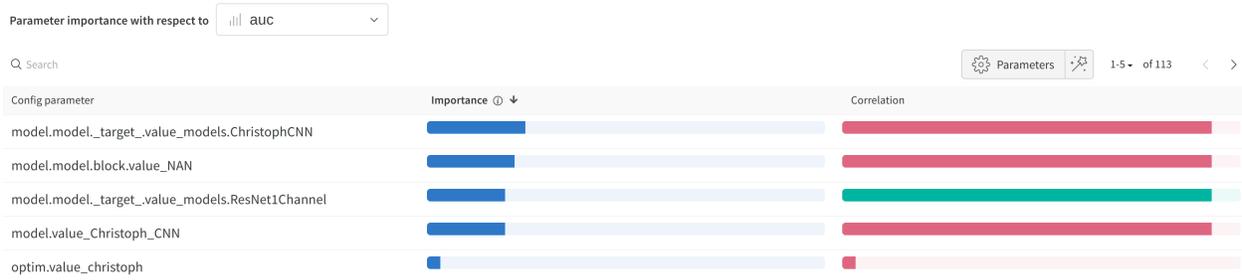

Figure 4: Parameter importance analysis for the hyper-parameter search sweep. The parameters are ordered by importance. A Positive correlation is indicated by a green color, while a negative correlation is indicated by a red color.

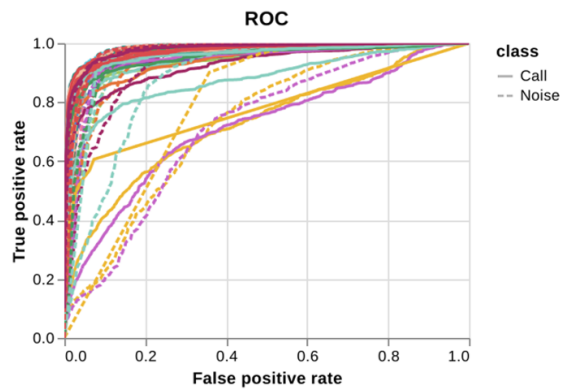

Figure 5: ROC curve of different runs in the hyper-parameters sweep experiment. Each run has a different color, dashed lines indicate the noise label, and continous lines indicate positive label.

### 3.3 Benchmark experiment

In the benchmark experiment we compare 3 different models used in the literature for marine mammals classification:

- ResNet50 model with Adam optimizer [55], which also serves as the default model in the Soundbay framework, and named "Default" in the table.
- ResNet50 model with SGD optimizer [36], and named "Jasco" in the table.
- A CNN suggested by Chin[59], named "Christoph" in the table. This model includes an optimization method suggested by the author, based on SGD optimizer.

The experiment consists of a grid run of all nine combinations of three models over three datasets. The first two models use Peak normalization as a preprocessing method, while the latter uses a sliding window preprocessing, as described in the method of Chin [59]. The three datasets are the ICML2013 dataset, Deep Voice's own Bazaruto 2018 recordings, and the OrcaData dataset. The commands to reproduce those experiments are given in 6.2 of Appendix A. Binary classification metrics can be found in table 2.

The default model performs best in terms of AUC on all datasets. The Jasco model has yielded limited results performing on the OrcaData dataset, while the Christoph model has produced limited results on the DeepVoice 2018 dataset. A major contributing factor to these results is the use of the SGD optimizer, which is less stable compared to the Adam optimizer [3] (the SGD optimizer is used both in the Jasco and Christoph models). Another evidence of this hypothesis is the negative correlation of using SGD optimizer with the AUC value, as observed in figure 4. It is essential to mention here that this experiment aims to demonstrate the ease of performing a benchmark analysis using Soundbay framework, and not necessarily find the best performing model. However, if needed, combining this benchmark with the hyper-parameters search capability can highly impact such an initiative.

| Model | Dataset | Accuracy | Precision | Recall | F1-score | AUC |
|---|---|---|---|---|---|---|
| Christopher | OrcaData | 0.936 | 0.849 | 0.603 | 0.705 | 0.902 |
| Jasco | OrcaData | 0.666 | 0.228 | 0.695 | 0.344 | 0.745 |
| Default | OrcaData | 0.893 | 0.547 | 0.877 | 0.67 | 0.915 |
| Chrisopher | ICML2013 | 0.951 | 0.847 | 0.486 | 0.618 | 0.878 |
| Jasco | ICML2013 | 0.958 | 0.732 | 0.761 | 0.746 | 0.958 |
| Default | ICML2013 | 0.974 | 0.938 | 0.728 | 0.82 | 0.984 |
| Chrisopher | DeepVoice 2018 | 0.41 | 1 | 0.009 | 0.6 | 0.769 |
| Jasco | DeepVoice 2018 | 0.632 | 0.84 | 0.47 | 0.6 | 0.769 |
| Default | DeepVoice 2018 | 0.906 | 0.875 | 0.981 | 0.925 | 0.966 |

Table 2: Binary classification metrics for the benchmark experiment.

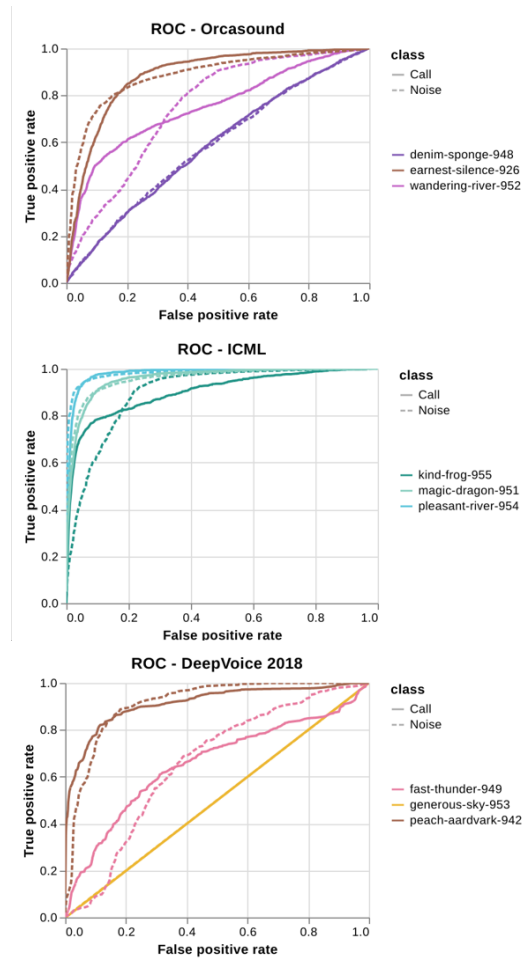

Figure 6: ROC curves of the 3 models (Christopher, Jasco and the Default model) on the different 3 datasets. Each dataset is in a different graph, and all three models are given per dataset.

# 4 Conclusion and future work

In the paper, we presented the Soundbay framework, a deep learning framework for bio-acoustic analysis. We discussed some of the main features provided by the framework, such as easy experiment launching, evaluation, and architecture search optimization. We further provided an essential baseline for comparing the performance of various models with regard to marine mammals' vocalizations detection, showing promising results for all the examined models. We hope this baseline will be adopted by the marine mammals' bioacoustic community. In the long run, this will aid in the development of better-automated vocalizations detection models. This will consequently lead to gaining bettrr4rr554rer insights with regard to marine mammals' communication.

In future work, we plan to regularly update the framework to include better performing models and state-of-the-art architecture, implementing an active-learning interface, and expanding the scope of framework experiments to a broader range of bio-acoustic applications, beyond marine mammal vocalizations. Deep Voice integrated the Soundbay framework into an end-to-end service, where we aim to provide bioacousticians with the tools required for efficient and reliable sound analysis.

We envision that Soundbay will serve as an assisting tool for many conservation applications, and will create value for decision-makers in the industry, academia, and government on the basis of using data to mitigate the trade-off between anthropogenic activites and wildlife conservation.

# 5 Acknowledgments

This research was supported by our dear social media followers and contributors to our crowd-funding campaign. We wish to thank the BCSS and its staff, including Dr. Mario Lebrato, Mrs. Karen Bowles, and Mr. Jason Morkel for their support in the data acquisition expedition in the Bazaruto archipelago. We wish to thank Dr. Oz Goffman and Mrs. Daphna Stern for taking part in the data acquisition expedition. Special thanks to Mr. Junio Fabrizio Borsani for his valuable tips and consultancy with regard to underwater acoustics. We wish to thank Mr. Liron Daniel for his dedicated annotations for the Mozambique 2018 dataset. We wish to thank Amazon AWS service for nonprofits, Microsoft for good and Deepchecks for the computing resources needed to carry out the experiments in this paper. Furthermore, we wish to thank Deepdub for their support and warm hosting while conducting this work. Finally, we wish to thank all of the Deep Voice members in the past and present for making this paper feasible.

# 6 Appendix A

## 6.1 Sweep Yaml

The following configuration was used to run the sweep of the hyper-parameters search described in 3.2:

```yaml
method: bayes
metric:
  goal: maximize
  name: auc
parameters:
  model:
    distribution: categorical
    values:
      - defaults
      - jasco_resnet_50
      - Christoph_CNN
  optim:
    distribution: categorical
    values:
      - defaults
      - christoph
  optim.epochs:
    value: 100
  data:
    value: icml2013
  preprocessors:
    distribution: categorical
    values:
      - _preprocessors
      - _preprocessors_sliding_window
      - _preprocessors_unit
  data.train_dataset.augmentations_p:
    distribution: uniform
    min: 0
    max: 1
  experiment.run_id:
    distribution: int_uniform
    min: 0
    max: 100000000
  experiment.group_name:
    value: "ICML Sweep"
early_terminate:
  type: hyperband
  min_iter: 10
program: train.py
command:
  - ${env}
  - ${interpreter}
  - ${program}
  - ${args_no_hyphens}
```

More details about sweeps configurations could be found in https://docs.wandb.ai/guides/sweeps/define-sweep-configuration

## 6.2 Experiments Reproducibility

All the runs configurations files for the experiments in the benchmark section 3.3 are provided in `soundbay/soundbay/conf/paper_experiments` Soundbay repository. Each configuration has a file name associated with the model and the dataset. For example, the file `christoph_icml.yaml` includes the following yaml:

```yaml
# @package _global_
defaults:
  - ../data: icml2013
  - ../augmentations: _augmentations
  - ../preprocessors: _preprocessors_sliding_window
  - ../model: Christoph_CNN
  - ../optim: christoph
  - ../experiment: defaults
```

The command to run this experiment:

```
python train.py -cn paper_experiments/christoph_icml
```

One can replace "christoph_icml" with any other experiment configuration from the paper.